\documentclass[a4paper,12pt]{iopart}
\usepackage{iopams}
\usepackage[dvips]{graphicx}
\usepackage{longtable}
\usepackage{multirow}
\usepackage{color}
\long\def\symbolfootnote[#1]#2{\begingroup%
\def\thefootnote{\fnsymbol{footnote}}\footnote[#1]{#2}\endgroup} 
\newcommand {\nc} {\newcommand}
\newcommand{\bsp}{\rule[0.5cm]{0.001cm}{0.0cm}}  
\nc {\beq} {\begin{eqnarray}}
\nc {\eeq} {\end{eqnarray}}
\nc {\eeqn} [1] {\label{#1} \end{eqnarray}}
\nc {\eoln} [1] {\label{#1} \\}
\nc {\eol} {\nonumber \\}
\nc {\la} {\mbox{$\langle$}}
\nc {\ra} {\mbox{$\rangle$}}
\nc {\cL} {\mbox{${\cal L}$}}
\nc {\dem} {\mbox{$\frac{1}{2}$}}
\nc {\ve} [1] {\mbox{\boldmath $#1$}}
\nc {\arrow} [2] {\mbox{$\mathop{\rightarrow}
\limits_{#1 \rightarrow #2}$}}
\begin{document}
\title{Quadrupole transitions in the bound rotational-vibrational spectrum 
of the deuterium molecular ion}
\author{Horacio Olivares Pil\'on}  
\address{Physique Quantique C.P. 165/82, and 
Physique Nucl\'eaire Th\'eorique et Physique Math\'ematique, C.P.\ 229, 
Universit\'e Libre de Bruxelles (ULB), B 1050 Brussels, Belgium}
\eads{\mailto{}}
\begin{abstract}
After the study of the three body molecular system H$_2^+$ ({\it J. Phys. B: At. Mol. Opt. Phys.} {\bf 45} 065101), its isotopomer, the deuterium molecular ion D$_2^+$ is studied. The three-body Schr\"odinger equation is solved using the Lagrange-mesh method in perimetric coordinates. Energies and wave functions for four vibrational states $v=0-3$ and bound or quasibound states for total orbital momenta from 0 to 56 are calculated. The 1986 fundamental constant  $m_d=3670.483014\,m_e$ is used. The obtained energies have an accuracy from about 13 digits for the lowest vibrational state to at least 9 digits for the third vibrational excited state. Quadrupole transition probabilities per time unit between those states over the whole rotational bands were calculated. Extensive results are presented with six significant figures. 
\end{abstract}
\submitto{}

\centerline{\today}
\maketitle
\section{Introduction}
\label{sec:intro}
Despite of the simplicity of the molecular ion H$_2^+$ and its isotopomers D$_2^+$ and T$_2^+$, the three body
Schr\"odinger equation can not be solved exactly. However, high precision results, for  energies 
and wave functions, have been obtained using different approaches. For the particular case of the deuterium molecular ion, the ground state as well as some vibrational states are well known.

In this sense, non-adiabatic dissociation energies were calculated for several vibrational-rotational levels for the ground electronic state~\cite{Mo93F} and also in the  first excited electronic state. In particular, the value of the ground state energy has been improved in several papers~\cite{BF02, YZL03,Mo99,HNN09}. Energies for the zero vibrational band for different values of the total angular momentum $L$ have been also considered~\cite{YZ04,HB03}.  Complete bound vibrational states below the dissociation limit $v\le 27$ were presented in~\cite{HBG00} for the two lowest rotational quantum numbers $L=0,1$. Extension to this work until $L=2$ is presented in~\cite{KH06} but with a different value of the deuteron mass.  Some expectation values have been also provided like the mean radii \cite{HB03,HNN09} or quadrupole moments \cite{HB03}. 

However, nowadays  quadrupole transitions has not been considered for this system. In this paper, the E2 transitions are obtained from the three-body wave functions calculated with the Lagrange-mesh method in perimetric coordinates \cite{HB99,HB01,HB03}, with which the calculation is particularly simple and very precise. The Lagrange-mesh method is an approximate variational calculation 
using a basis of Lagrange functions and the associated Gauss quadrature. 
It has the high accuracy of a variational approximation and the simplicity of a calculation 
on a mesh. 

In section \ref{sec:lmtp}, some expressions are presented but detailed explanations are given in~\cite{OB12T}.
In section \ref{sec:res}, energies are given for the lowest four vibrational 
states over the full rotational bands and E2 transition probabilities are tabulated. 
Conclusions are presented in section \ref{sec:conc}. Throughout atomic units are used. 

%
\section{Lagrange-mesh calculation of transition probabilities}
\label{sec:lmtp}
Quadrupole oscillator strength for an electric transition between an initial state $i$ and a final state $f$ is given by~\cite{So72,So79} 
\beq
f_{i\rightarrow f}^{(2)} = \frac{1}{30}\alpha^{2} \left(E_f - E_i\right)^{3} \frac{S_{if}^{(2)}}{2J_i+1},
\eeqn{2.6}
where $\alpha$ is the fine-structure constant and 
\beq
S_{if}^{(2)} = S_{fi}^{(2)} 
= \sum_{M_iM_f\mu} |\la \gamma_i J_i M_i| Q^{(2)}_\mu | \gamma_f J_f M_f \ra |^2 
= |\la \gamma_i J_i || Q^{(2)} || \gamma_f J_f \ra |^2,
\eeqn{2.3}
is the reduced matrix element, defined according to \cite{Ed57}. Here, $J_{i,f}$ is a total angular momentum, $M_{i,f}$ is its projection and $\gamma_{i,f}$ represents the other quantum numbers. The transition probability per time unit for $E_f < E_i$ is given (the atomic unit of time is $a_0/\alpha c \approx 2.4188843 \times 10^{-17}$ s) by 
\beq
W_{i\rightarrow f}^{(2)} = \frac{1}{15} \alpha^{5}(E_i - E_f)^{5} \frac{S_{if}^{(2)}}{2J_i+1}\,.
\eeqn{2.8}

After the elimination of the center-of-mass motion, the system can be described using the Euler angles $(\psi,\phi,\phi)$, defining the orientation of the triangle formed by the three particles and three internal coordinates describing the shape of the triangle. The internal degrees of freedom are described using the perimetric coordinates ($x,y,z$) defined as
\beq
\begin{array}{rcl}
x &=& R-r_{e2}+r_{e1}, \\
y &=& R+r_{e2}-r_{e1}, \\
z &=& -R+r_{e2}+r_{e1},
\end{array}
\eeqn{3.1}
where $r_{e1}$ and $r_{e2}$ are the distances between the electron and deuterons 1 and 2, respectively. 
The domains of variation of these six variables are $[0,2\pi]$ for $\psi$ and $\phi$, $[0,\pi]$ for $\theta$ and $[0,\infty[$ for $x, y$ and $z$. In perimetric coordinates, the transition irreducible tensor operator $Q^{(2)}_\mu$ is given  by
\beq \fl
Q^{(2)}_\mu = \frac{1}{2} [R^2 - \gamma (2\zeta^2 - \rho^2)] D^{2}_{\mu 0}(\psi,\theta,\phi) 
- \sqrt{\frac{3}{2}} \gamma \zeta \rho [D^{2}_{\mu 1}(\psi,\theta,\phi) - D^{2}_{\mu -1}(\psi,\theta,\phi)]
\eol
- \sqrt{\frac{3}{8}} \gamma \rho^2 [D^{2}_{\mu 2}(\psi,\theta,\phi) + D^{2}_{\mu -2}(\psi,\theta,\phi)].
\eeqn{3.4}
where 
\beq
\gamma = 1 - \frac{2m_e}{M} - \frac{m_e^2}{M^2}
\eeqn{3.5}
and the expression of $R$, $\rho$ and $\zeta$ in terms of perimetric coordinates are given for example in~\cite{HB01,OB12T}.

In order to evaluate~(\ref{2.3}) it is necessary to solve three-body Hamiltonian that  involves Coulomb forces between the particles but no spin-dependent forces. The total orbital momentum $L$ and parity $\pi$ are constants of motion. The wave functions with orbital momentum $L$ and parity $\pi$ are expanded as \cite{HB03} 
\beq
\Psi^{(L^{\pi})\sigma}_M = \sum_{K=0}^L {\cal D}_{MK}^{L\pi}(\psi,\theta,\phi) \Phi_K^{(L^{\pi})\sigma}(x,y,z). 
\eeqn{4.1}
In practice, the sum can be truncated at some value $K_{\rm max}$. 
The normalized angular functions ${\cal D}_{MK}^{L\pi}(\psi,\theta,\phi)$ are defined for $K \ge 0$ by
\beq
{\cal D}_{MK}^{L\pi}(\psi,\theta,\phi) &=& \frac{\sqrt{2L+1}}{4\pi} \left(1+\delta_{K0}\right)^{-1/2}
\left[D_{MK}^L(\psi,\theta,\phi) \right.\nonumber\\
&& \left.+ \pi (-1)^{L+K} D_{M\;-K}^L(\psi,\theta,\phi)\right],
\eeqn{4.2}
where $D_{MK}^L(\psi,\theta,\phi)$ represents a Wigner matrix element. 
They have parity $\pi$ and change as $\pi(-1)^K$ under permutation of the deuterons. 
The $\Phi_K^{(L^{\pi})\sigma}$ functions with symmetry $\sigma$  are symmetric for $(-1)^K = \sigma\pi$ and antisymmetric for 
$(-1)^K = -\sigma\pi$, when $x$ and $y$ are exchanged and are expanded in the Lagrange basis as
\beq
\Phi_K^{(L^{\pi})\sigma}(x,y,z) &=& \sum_{i=1}^{N} \sum_{j=1}^{i-\delta_K} \sum_{k=1}^{N_z} 
C_{Kijk}^{(L^{\pi})\sigma} \left[ 2 (1+\delta_{ij})\right]^{-1/2}\nonumber\\
&& \times \left[ F^K_{ijk}(x,y,z) + \sigma \pi (-1)^K F^K_{jik}(x,y,z) \right],
\eeqn{5.5}
where $N$, $N_z$ are the number of points in the $x$ and $z$. Scale factors $h$ and $h_z$ are introduced in order to fit the meshes to the size of the physical problem (the same number $N$ of mesh points and the same scale factor $h$ are used for the two perimetric coordinates $x$ and $y$). Because of the symmetrization the sum over $j$ is limited by the value $i-\delta_K$, where $\delta_K$ is equal to 0 when $(-1)^K=\sigma \pi$ and to 1 when $(-1)^K=-\sigma \pi$.

The three-body Hamiltonian in perimetric coordinates for each good quantum number $L$ 
and its discretization on a Lagrange mesh are given in \cite{HB01}.  For given $L^\pi$, the eigenvalues in increasing order are labeled by a quantum number $v \ge 0$ 
related to the vibrational excitation in the Born-Oppenheimer picture. 
The corresponding eigenvectors provide the coefficients appearing in expansion \eref{5.5}. Because the $K$-value appears in the eigenfunction~(\ref{4.1}), the matrix element~(\ref{2.3}) contains terms where $K_f=K_i$, $K_f=K_i\pm 1$ and $K_f=K_i\pm 2$. These terms are labeled as $\kappa = 0,1,2$, respectively (see~\cite{OB12T}).   
 
\section{Energies and quadrupole transition probabilities}
\label{sec:res}
Using the Lagrange-mesh method, energies of the $v = 0$ lowest vibrational bound states for some values of the total angular momentum from $L=0$ to 51 have been calculated in~\cite{HB03}. At the present study, 
those calculations are extended for all bound  and quasibound states up to $L = 56$ and four vibrational states $v=0-3$. In order to calculate transition matrix elements 
involving two different wave functions, a single three-dimensional mesh is used for all the vibro-rotational states in consideration: $N = N_x = N_y = 45$ and $N_z = 20$. A single set of scaling factors are also employed: $h = h_x = h_y = 0.08$, $h_z = 0.4$. For a given $K$ value, the total number of basis states is 20700 or 19800 depending on the parity of $K$. For $K>2$, calculations are performed with $K_{\rm max} = 2$. 
They correspond to a size of 61200. Here,  the 1986 fundamental constant  $m_d=3670.483014\,m_e$ is  adopted and the dissociation energy is $E_d=-0.499\,863\,815\,249$.

Table \ref{tab:1} presents the energies obtained for the three lowest vibrational states $v=0-3$ as a function of the rotational quantum number. The accuracy is estimated from the stability of the digits with respect to calculations with $N \pm 2$ mesh points. The error is expected to be at most of a few units on the last displayed digit. Comparisons when are available are displayed below at the same line for each $L$-value. 

\begin{center}
\begin{longtable}{rllll}
\caption{Energies of the four lowest vibrational bound or quasibound states 
in the $\Sigma_g$ rotational band of the D$_2^+$ molecular ion. 
Quasibound states are separated from bound states by a horizontal bar. 
For each $L$ value, the Lagrange-mesh results obtained with $N_x = N_y = 45$, $N_z = 20$ and 
$h_x = h_y = 0.08$, $h_z = 0.4$ are presented in the first line. 
Other references are indicated ($^a$:\cite{HBG00}, $^b$:\cite{Mo99}, 
, $^c$: \cite{YZ04}, $^d$: \cite{Mo93F}). 
The deuteron mass is taken as $m_d=3670.483014\,m_e$.}
\label{tab:1}\\
\hline 
$L$& $v=0$          & $v=1$           &  $v=2$          &$v=3$\\
\hline
\endfirsthead
\multicolumn{5}{c}{{\tablename} \thetable{} -- Continuation}\\ 
\hline
$L$& $v=0$          & $v=1$           &  $v=2$          &$v=3$\\
\hline
\endhead
\hline
\multicolumn{5}{l}{{Continued on Next Page\ldots}}\\
\endfoot
\hline                                       
\endlastfoot
 0& -0.598\,788\,784\,330\,7& -0.591\,603\,121\,903& -0.584\,712\,207\,01&  -0.578\,108\,591\,4\\ 
  & -0.598\,788\,784\,330\,68$^{a}$&-0.591\,603\,121\,903\,21$^{a}$
  & -0.584\,712\,207\,009\,99$^{a}$&-0.578\,108\,591\,436\,87$^{a}$\\
  & -0.598\,788\,784\,330\,7$^{b}$&-0.591\,603\,121\,903\,2[4]$^{b}$&---&---\\ 
  & -0.598\,788\,784\,32$^{d}$& -0.591\,603\,121\,92$^{d}$
  & -0.584\,712\,206\,99$^{d}$& -0.578\,108\,591\,41$^{d}$\\
 \bsp
 1& -0.598\,654\,873\,220\,5& -0.591\,474\,211\,529& -0.584\,588\,169\,62&  -0.577\,989\,312\,0\\ 
  & -0.598\,654\,873\,22$^{a}$& -0.591\,474\,211\,53$^{a}$
  & -0.584\,588\,169\,62$^{a}$&-0.577\,989\,311\,96$^{a}$\\
  & -0.598\,654\,873\,220\,5$^{b}$& -0.591\,474\,211\,528\,6[7]$^{b}$&---&---\\ 
  & -0.598\,654\,873\,21$^{d}$& -0.591\,474\,211\,53$^{d}$
  & -0.584\,588\,169\,60$^{d}$& -0.577\,989\,311\,98$^{d}$\\
 \bsp
 2& -0.598\,387\,585\,810\,0& -0.591\,216\,909\,624& -0.584\,340\,598\,38&  -0.577\,751\,241\,9\\ 
  & -0.598\,387\,585\,809\,98$^{c}$&&&\\
  & -0.598\,387\,585\,78$^{d}$& -0.591\,216\,909\,63$^{d}$
  & -0.584\,340\,598\,39$^{d}$& -0.577\,751\,241\,91$^{d}$\\
 \bsp
 3& -0.597\,987\,984\,746\,7& -0.590\,832\,247\,071& -0.583\,970\,493\,74&  -0.577\,395\,352\,5\\ 
  & -0.597\,987\,984\,746\,72$^{c}$&&&\\
  & -0.597\,987\,984\,75$^{d}$& -0.590\,832\,247\,09$^{d}$
  & -0.583\,970\,493\,75$^{d}$& -0.577\,395\,352\,53$^{d}$\\
 \bsp
 4& -0.597\,457\,646\,783\,8& -0.590\,321\,753\,376& -0.583\,479\,339\,94&  -0.576\,923\,084\,8\\ 
  & -0.597\,457\,646\,783\,78$^{c}$&&&\\
  & -0.597\,457\,646\,78$^{d}$& -0.590\,321\,753\,37$^{d}$
  & -0.583\,479\,339\,94$^{d}$& -0.576\,923\,084\,78$^{d}$\\
  \bsp
 5& -0.596\,798\,642\,700\,8& -0.589\,687\,437\,079& -0.582\,869\,085\,94&  -0.576\,336\,330\,6\\ 
  & -0.596\,798\,642\,700\,81$^{c}$&&&\\
  & -0.596\,798\,642\,68$^{d}$& -0.589\,687\,437\,09$^{d}$
  & -0.582\,869\,085\,95$^{d}$& -0.576\,336\,330\,45$^{d}$\\  
  \bsp
 6& -0.596\,013\,511\,469\,5& -0.588\,931\,760\,548& -0.582\,142\,120\,82&  -0.575\,637\,408\,9\\ 
  & -0.596\,013\,511\,469\,53$^{c}$&&&\\
 7& -0.595\,105\,229\,402\,0& -0.588\,057\,609\,916& -0.581\,301\,244\,47&  -0.574\,829\,037\,2\\ 
  & -0.595\,105\,229\,401\,95$^{c}$&&&\\
 8& -0.594\,077\,175\,129\,3& -0.587\,068\,260\,974& -0.580\,349\,634\,45&  -0.573\,914\,299\,3\\ 
  & -0.594\,077\,175\,129\,31$^{c}$&&&\\
 9& -0.592\,933\,091\,328\,5& -0.585\,967\,341\,945& -0.579\,290\,809\,74&  -0.572\,896\,610\,0\\ 
  & -0.592\,933\,091\,328\,53$^{c}$&&&\\
10& -0.591\,677\,044\,141\,4& -0.584\,758\,794\,049& -0.578\,128\,592\,47&  -0.571\,779\,678\,0\\ 
  & -0.591\,677\,044\,141\,33$^{c}$&&&\\
11& -0.590\,313\,381\,220\,4& -0.583\,446\,830\,801& -0.576\,867\,068\,41&  -0.570\,567\,467\,4\\ 
  & -0.590\,313\,381\,220\,32$^{c}$&&&\\
12& -0.588\,846\,689\,292\,3& -0.582\,035\,896\,884& -0.575\,510\,547\,15&  -0.569\,264\,159\,6\\ 
  & -0.588\,846\,689\,292\,25$^{c}$&&&\\
13& -0.587\,281\,752\,056\,3& -0.580\,530\,627\,415& -0.574\,063\,522\,70&  -0.567\,874\,114\,7\\ 
14& -0.585\,623\,509\,141\,9& -0.578\,935\,808\,297& -0.572\,530\,635\,17&  -0.566\,401\,834\,8\\ 
15& -0.583\,877\,016\,744\,9& -0.577\,256\,338\,258& -0.570\,916\,634\,23&  -0.564\,851\,928\,8\\ 
16& -0.582\,047\,410\,445\,6& -0.575\,497\,193\,053& -0.569\,226\,344\,58&  -0.563\,229\,078\,7\\ 
17& -0.580\,139\,870\,600\,8& -0.573\,663\,392\,212& -0.567\,464\,634\,07&  -0.561\,538\,010\,0\\ 
18& -0.578\,159\,590\,589\,4& -0.571\,759\,968\,586& -0.565\,636\,384\,50&  -0.559\,783\,462\,5\\ 
19& -0.576\,111\,748\,091\,6& -0.569\,791\,940\,871& -0.563\,746\,465\,33&  -0.557\,970\,166\,4\\ 
20& -0.574\,001\,479\,489\,6& -0.567\,764\,289\,167& -0.561\,799\,710\,40&  -0.556\,102\,819\,0\\ 
21& -0.571\,833\,857\,401\,5& -0.565\,681\,933\,603& -0.559\,800\,897\,57&  -0.554\,186\,066\,4\\ 
22& -0.569\,613\,871\,292\,8& -0.563\,549\,715\,941& -0.557\,754\,731\,37&  -0.552\,224\,486\,3\\ 
23& -0.567\,346\,411\,059\,8& -0.561\,372\,384\,075& -0.555\,665\,828\,34&  -0.550\,222\,575\,0\\ 
24& -0.565\,036\,253\,438\,8& -0.559\,154\,579\,261& -0.553\,538\,705\,16&  -0.548\,184\,735\,6\\ 
25& -0.562\,688\,051\,068\,4& -0.556\,900\,825\,932& -0.551\,377\,769\,16&  -0.546\,115\,270\,0\\ 
26& -0.560\,306\,324\,011\,5& -0.554\,615\,523\,892& -0.549\,187\,311\,26&  -0.544\,018\,372\,2\\ 
27& -0.557\,895\,453\,538\,2& -0.552\,302\,942\,710& -0.546\,971\,501\,01&  -0.541\,898\,124\,2\\ 
28& -0.555\,459\,677\,964\,5& -0.549\,967\,218\,120& -0.544\,734\,383\,64&  -0.539\,758\,493\,9\\ 
29& -0.553\,003\,090\,349\,8& -0.547\,612\,350\,239& -0.542\,479\,878\,85&  -0.537\,603\,334\,2\\ 
30& -0.550\,529\,637\,862\,4& -0.545\,242\,203\,435& -0.540\,211\,781\,30&  -0.535\,436\,385\,0\\ 
31& -0.548\,043\,122\,635\,1& -0.542\,860\,507\,670& -0.537\,933\,762\,62&  -0.533\,261\,275\,0\\ 
32& -0.545\,547\,203\,949\,6& -0.540\,470\,861\,200& -0.535\,649\,374\,77&  -0.531\,081\,527\,3\\ 
33& -0.543\,045\,401\,606\,3& -0.538\,076\,734\,487& -0.533\,362\,054\,69&  -0.528\,900\,563\,8\\ 
34& -0.540\,541\,100\,352\,2& -0.535\,681\,475\,235& -0.531\,075\,130\,31&  -0.526\,721\,713\,7\\ 
35& -0.538\,037\,555\,267\,4& -0.533\,288\,314\,477& -0.528\,791\,827\,57&  -0.524\,548\,221\,2\\ 
36& -0.535\,537\,898\,025\,9& -0.530\,900\,373\,673& -0.526\,515\,278\,89&  -0.522\,383\,256\,3\\ 
37& -0.533\,045\,143\,977\,9& -0.528\,520\,672\,798& -0.524\,248\,532\,76&  -0.520\,229\,927\,0\\ 
38& -0.530\,562\,200\,023\,2& -0.526\,152\,139\,463& -0.521\,994\,564\,82&  -0.518\,091\,292\,9\\ 
39& -0.528\,091\,873\,277\,5& -0.523\,797\,619\,127& -0.519\,756\,290\,58&  -0.515\,970\,382\,5\\ 
40& -0.525\,636\,880\,568\,4& -0.521\,459\,886\,554& -0.517\,536\,580\,00&  -0.513\,870\,212\,6\\ 
41& -0.523\,199\,858\,840\,0& -0.519\,141\,658\,708& -0.515\,338\,274\,50&  -0.511\,793\,812\,4\\ 
42& -0.520\,783\,376\,595\,1& -0.516\,845\,609\,396& -0.513\,164\,206\,99&  -0.509\,744\,252\,7\\ 
43& -0.518\,389\,946\,571\,7& -0.514\,574\,386\,104& -0.511\,017\,225\,87&  -0.507\,724\,683\,2\\ 
44& -0.516\,022\,039\,937\,6& -0.512\,330\,629\,670& -0.508\,900\,224\,43&  -0.505\,738\,379\,5\\ 
45& -0.513\,682\,102\,404\,5& -0.510\,116\,997\,691& -0.506\,816\,177\,71&  -0.503\,788\,806\,7\\ 
46& -0.511\,372\,572\,833\,1& -0.507\,936\,193\,030& -0.504\,768\,189\,97&  -0.501\,879\,706\,3\\ 
47& -0.509\,095\,905\,143\,5& -0.505\,790\,999\,399& -0.502\,759\,557\,97&  -0.500\,015\,221\,0\\ \cline{5-5}
48& -0.506\,854\,594\,707\,7& -0.503\,684\,327\,110& -0.500\,793\,858\,28&  -0.498\,200\,082\,9\\ \cline{4-4}
49& -0.504\,651\,210\,969\,6& -0.501\,619\,273\,844& -0.498\,875\,073\,34&  -0.496\,439\,916\,9\\ \cline{3-3}
50& -0.502\,488\,438\,936\,8& -0.499\,599\,208\,500& -0.497\,007\,783\,09&  -0.494\,741\,77    \\ 
51& -0.500\,369\,133\,705\,2& -0.497\,627\,892\,183& -0.495\,197\,476\,73&  -0.493\,115\,1     \\ \cline{2-2} 
52& -0.498\,296\,394\,825\,7& -0.495\,709\,662\,497& -0.493\,451\,107    &  \\
53& -0.496\,273\,672\,262\,6& -0.493\,849\,734\,3  & -0.491\,778\,2      &  \\ 
54& -0.494\,304\,925\,576\,7& -0.492\,054\,739     & &\\ 
55& -0.492\,394\,879\,740   & -0.490\,333\,83      & &\\
56& -0.490\,549\,476        & &&\\
\end{longtable}
\end{center}             
The ground state $(L^\pi,v) = (0^+,0)$ is well known and its value has been improved in several papers~\cite{YZL03,HBG00,BF02}. In particular Hijikata $et.al.$~\cite{HNN09} reach an accuracy of 27 digits. In this sense we have an accuracy about $10^{-13}$. 

Energies of the $v=0$ vibrational band for $L=2-12$ were presented by Zong $et.al.$~\cite{YZ04}. Moss~\cite{Mo93F} presents not only non-adiabatic dissociation-energies but also relativistic and radiative corrections for bound vibrational states up to $L = 5$. Seven rovibrational states of the first electronic state are also presented there. More precise results for all the vibrational levels below the dissociation limit $v\le 27$ are given by Hilico $et.al.$~\cite{HBG00} for the two lowest rotational quantum numbers $L=0,1$ . Beyond that, Karr $et.al.$~\cite{KH06} extended this results until $L=2$ but with a different value of the deuteron mass $m_d=3670.4829652$. In order to make some comparison  with those results, Lagrange-mesh calculations considering this value of the deuteron mass are given in Table~\ref{tab:1b}. Although less digits were obtained when comparing with~\cite{KH06} we can consider that the agreement is in all the digits presented.
 
\begin{figure}[hbt]
\setlength{\unitlength}{1mm}
\begin{picture}(140,70) (-20,10) 
\put(0,0){\mbox{\scalebox{1.5}{\includegraphics{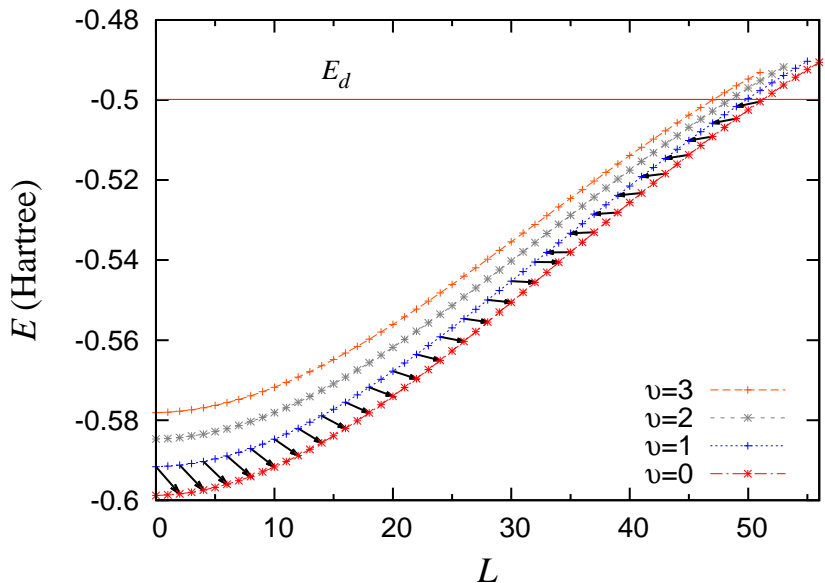}}}}
\end{picture} \\
\caption{Four lowest $\Sigma_g$ rotational bands of the D$_2^+$ molecular ion 
and dissociation energy $E_d$. Arrows show how the direction of $L \rightarrow L+2$
transitions between the two lowest bands changes along the band.}
\label{fig:1}
\end{figure}

The accuracies for the first, second and third rotational bands are $10^{-11}$, $10^{-10}$ and $10^{-9}$, 
respectively. The major comparison is possible  with  the results given by Moss \cite{Mo93F}. The obtained spectrum is presented in Fig.~\ref{fig:1}. 

\begin{table}[hbt]
\centering{
\caption{Comparison of the total energy for three rotational states $L=0-2$
and four vibrational states $v=0-3$ taking $m_d=3670.4829652$. For each value of the 
rotational quantum number, the first line is calculated using the Lagrange mesh 
method with  $N=45$ and $N_z=20$ and the scale factors are
$h=0.08$ and $h_z=0.4$. The second line are the results presented by Karr $et.al.$\cite{KH06}.} 
\label{tab:1b} 
\resizebox{16cm}{!}{
\begin{tabular}{lllllll}
\hline
$L$& $v=0$          & $v=1$           &  $v=2$          &$v=3$\\
\hline
0&-0.598\,788\,784\,304\,46 &-0.591\,603\,121\,831\,2  &-0.584\,712\,206\,896     &-0.578\,108\,591\,28   \\
 &-0.598\,788\,784\,304\,46 &-0.591\,603\,121\,831\,23 &-0.584\,712\,206\,896\,08 &-0.578\,108\,591\,28475\\
1&-0.598\,654\,873\,192\,5  &-0.591\,474\,211\,454\,9  &-0.584\,588\,169\,503     &-0.577\,989\,311\,80   \\ 
 &-0.598\,654\,873\,192\,49 &-0.591\,474\,211\,454\,95 &-0.584\,588\,169\,503\,36 &-0.577\,989\,311\,80781\\ 
2&-0.598\,387\,585\,778\,5  &-0.591\,216\,909\,547\,5  &-0.584\,340\,598\,262     &-0.577\,751\,241\,73   \\ 
 &-0.598\,387\,585\,778\,48 &-0.591\,216\,909\,547\,45 &-0.584\,340\,598\,262\,38 &-0.577\,751\,241\,73967\\  
\hline
\end{tabular}
}}
\end{table}

\begin{table}[hbt]
\centering{
\caption{Convergence of the energies and transition probabilities 
as a function of the numbers $N$ and $N_z$ of mesh points. 
Two cases are shown: $(4^+,0) \rightarrow (2^+,0)$ where $L_f = L_i - 2$ (upper set) 
and $(43^+,3) \rightarrow (45^+,1)$ where $L_f = L_i + 2$ (lower set). 
The scale factors are $h = 0.08$ and $h_z = 0.4$.} 
\label{tab:2} 
\resizebox{16cm}{!}{
\begin{tabular}{lllllll}
\hline
$N$&$N_z$&$E_i (4^+,0)$&$E_f (2^+,0)$&$W_0$ ($10^{-11}$s$^{-1}$)&$W_1$ ($10^{-11}$s$^{-1}$)&$W$ ($10^{-11}$s$^{-1}$)\\
\hline
20&20&-0.597\,449\,5           &-0.598\,379\,6           &2.97           &2.97           &2.97       \\
30&20&-0.597\,457\,646\,20     &-0.598\,387\,585\,24     &2.966\,273     &2.965\,657\,294&2.965\,657 \\
30&30&-0.597\,457\,646\,20     &-0.598\,387\,585\,24     &2.966\,273     &2.965\,657\,295&2.965\,657 \\
40&20&-0.597\,457\,646\,783\,75&-0.598\,387\,585\,809\,95&2.966\,271\,937&2.965\,656\,618&2.965\,656\,329\\
40&30&-0.597\,457\,646\,783\,76&-0.598\,387\,585\,809\,97&2.966\,271\,938&2.965\,656\,618&2.965\,656\,329\\
45&20&-0.597\,457\,646\,783\,80&-0.598\,387\,585\,809\,99&2.966\,271\,937&2.965\,656\,618&2.965\,656\,329\\
50&20&-0.597\,457\,646\,783\,80&-0.598\,387\,585\,809\,99&2.966\,271\,937&2.965\,656\,618&2.965\,656\,329\\
\hline
$N$&$N_z$&$E_i (43^+,2)$&$E_f (45^+,0)$&$W_0$ ($10^{-11}$s$^{-1}$)&$W_1$ ($10^{-11}$s$^{-1}$)&$W$ ($10^{-11}$s$^{-1}$)\\
\hline
30&20&-0.507\,54           &-0.510\,115\,1           &7.966\,532     &7.965\,738     &7.965\,752  \\
30&30&-0.507\,54           &-0.510\,115\,1           &7.966\,532     &7.965\,738     &7.965\,752  \\
40&20&-0.507\,724\,681\,4  &-0.510\,116\,997\,683    &5.356\,451\,153&5.355\,922\,250&5.355\,931\,775\\
40&30&-0.507\,724\,681\,4  &-0.510\,116\,997\,683    &5.356\,451\,152&5.355\,922\,249&5.355\,931\,775\\
45&20&-0.507\,724\,683\,174&-0.510\,116\,997\,691\,28&5.356\,445\,629&5.355\,916\,726&5.355\,926\,252\\
50&20&-0.507\,724\,683\,184&-0.510\,116\,997\,691\,33&5.356\,445\,608&5.355\,916\,705&5.355\,926\,230\\
\hline
\end{tabular}
}}
\end{table}
Two cases of convergence tests are displayed in Table \ref{tab:2}. For the values $h = 0.14$ and $h_z = 0.4$, initial and final states are shown as well as the transition probability per second for different values of $N$ and $N_z$. The transition probability $W_0$ is obtained by restricting the matrix element~(\ref{2.3}) to the case where  $\kappa = 0$ while $W_1$ corresponds to $\kappa \le 1$. The $\kappa = 1$ contributions have an importance smaller than 0.05 \% 
and that the $\kappa = 2$ contributions are smaller than 0.005 \%. A good convergence with respect to $N_z$ is already obtained for $N_z = 20$. The convergence with respect to $N$ is slower.  
Since the convergence is exponential, one can estimate that 
the relative accuracy on $W$ is about $10^{-9}$ for $(4^+,0)\rightarrow(2^+,0)$ 
and still better than $10^{-8}$ for $(43^-,2)\rightarrow(35^-,0)$. 
Similar tests have been performed for other transitions. 

The convergence of the transition probabilities with respect to $K_{\rm max}$ can be studied 
by comparison with results from wave functions truncated at $K_{\rm max} = 0$ 
and $K_{\rm max}= 1$. 
The relative error when $K_{\rm max} = 0$ is smaller than 0.3  for all considered transitions 
while the error for $K_{\rm max} = 1$ is smaller than $10^{-5}$. 
This is rather similar to truncations with respect to $\kappa$. 
By extrapolation, we estimate that the relative error on the present transition probabilities 
obtained with $K_{\rm max}= 2$ should be smaller than $10^{-7}$. The number of digits presented in Tables~\ref{tab:3} and \ref{tab:4} are the same if we consider the deuteron mass as $m_d=3670.482962$.

Table \ref{tab:3} presents the transitions probabilities per second for transitions within a same rotational band, $L_f = L_i - 2$ and $v_f = v_i \le 3$. They include some transition probabilities involving quasibound states. Six significant digits are given. These probabilities increase slowly with $L$ with a maximum around $L_i = 45$, $43$, $41$ and $40$  for $v_i = 0$, $1$, $2$ and $3$, respectively, which is due to a maximum of the energy differences around $L_i = 35$. 

\begin{center}
\begin{longtable}{rllll}
\caption{Quadrupole transition probabilities per second $W$ for transitions 
between states of a same rotational band ($v_f = v_i$, $L_f = L_i - 2$). 
Results are given with five digits followed by the power of 10. } 
\label{tab:3} \\
\\[-4.9ex]
\hline
$L_i$&$v_i=0$&$v_i=1$&$v_i=2$&$v_i=3$\\
\hline
\endfirsthead
\multicolumn{4}{c}{{\tablename} \thetable{} -- Continuation}\\
\hline
$L_i$&$v_i=0$&$v_i=1$&$v_i=2$&$v_i=3$\\
\hline
\endhead
\hline
\multicolumn{4}{l}{{Continued on Next Page\ldots}}\\
\endfoot
\hline
\endlastfoot
 2& 3.066\,49-13& 3.069\,47-13& 3.041\,09-13& 2.983\,89-13\\ 
 3& 5.026\,89-12& 5.028\,74-12& 4.979\,40-12& 4.883\,07-12\\ 
 4& 2.965\,66-11& 2.964\,08-11& 2.932\,50-11& 2.873\,43-11\\ 
 5& 1.086\,42-10& 1.084\,55-10& 1.071\,79-10& 1.049\,07-10\\ 
 6& 3.012\,58-10& 3.003\,01-10& 2.963\,57-10& 2.896\,89-10\\ 
 7& 6.952\,82-10& 6.918\,74-10& 6.816\,71-10& 6.652\,88-10\\ 
 8& 1.408\,15-09& 1.398\,48-09& 1.375\,27-09& 1.339\,80-09\\ 
 9& 2.585\,58-09& 2.562\,15-09& 2.514\,33-09& 2.444\,51-09\\ 
10& 4.397\,74-09& 4.347\,29-09& 4.256\,29-09& 4.128\,87-09\\ 
11& 7.033\,38-09& 6.934\,41-09& 6.772\,21-09& 6.553\,53-09\\ 
12& 1.069\,24-08& 1.051\,23-08& 1.023\,87-08& 9.882\,24-09\\ 
13& 1.557\,73-08& 1.526\,92-08& 1.482\,93-08& 1.427\,32-08\\ 
14& 2.188\,42-08& 2.138\,42-08& 2.070\,56-08& 1.987\,05-08\\ 
15& 2.979\,43-08& 2.901\,83-08& 2.800\,88-08& 2.679\,62-08\\ 
16& 3.946\,54-08& 3.830\,72-08& 3.685\,31-08& 3.514\,39-08\\ 
17& 5.102\,50-08& 4.935\,41-08& 4.731\,91-08& 4.497\,32-08\\ 
18& 6.456\,45-08& 6.222\,54-08& 5.945\,02-08& 5.630\,65-08\\ 
19& 8.013\,50-08& 7.694\,71-08& 7.324\,96-08& 6.912\,69-08\\ 
20& 9.774\,47-08& 9.350\,26-08& 8.867\,95-08& 8.337\,83-08\\ 
21& 1.173\,58-07& 1.118\,33-07& 1.056\,62-07& 9.896\,63-08\\ 
22& 1.388\,95-07& 1.318\,39-07& 1.240\,79-07& 1.157\,61-07\\ 
23& 1.622\,35-07& 1.533\,82-07& 1.437\,81-07& 1.336\,00-07\\ 
24& 1.872\,20-07& 1.762\,88-07& 1.645\,82-07& 1.522\,94-07\\ 
25& 2.136\,53-07& 2.003\,54-07& 1.862\,73-07& 1.716\,29-07\\ 
26& 2.413\,12-07& 2.253\,47-07& 2.086\,20-07& 1.913\,74-07\\ 
27& 2.699\,45-07& 2.510\,18-07& 2.313\,75-07& 2.112\,85-07\\ 
28& 2.992\,83-07& 2.770\,98-07& 2.542\,75-07& 2.311\,09-07\\ 
29& 3.290\,40-07& 3.033\,08-07& 2.770\,52-07& 2.505\,89-07\\ 
30& 3.589\,21-07& 3.293\,66-07& 2.994\,36-07& 2.694\,71-07\\ 
31& 3.886\,27-07& 3.549\,86-07& 3.211\,56-07& 2.875\,02-07\\ 
32& 4.178\,58-07& 3.798\,85-07& 3.419\,51-07& 3.044\,40-07\\ 
33& 4.463\,17-07& 4.037\,86-07& 3.615\,64-07& 3.200\,52-07\\ 
34& 4.737\,16-07& 4.264\,25-07& 3.797\,53-07& 3.341\,19-07\\ 
35& 4.997\,78-07& 4.475\,48-07& 3.962\,89-07& 3.464\,37-07\\ 
36& 5.242\,38-07& 4.669\,15-07& 4.109\,58-07& 3.568\,17-07\\ 
37& 5.468\,49-07& 4.843\,05-07& 4.235\,64-07& 3.650\,90-07\\ 
38& 5.673\,81-07& 4.995\,14-07& 4.339\,29-07& 3.711\,01-07\\ 
39& 5.856\,21-07& 5.123\,55-07& 4.418\,90-07& 3.747\,15-07\\ 
40& 6.013\,78-07& 5.226\,61-07& 4.473\,06-07& 3.758\,14-07\\ 
41& 6.144\,80-07& 5.302\,84-07& 4.500\,52-07& 3.742\,96-07\\ 
42& 6.247\,75-07& 5.350\,96-07& 4.500\,19-07& 3.700\,73-07\\ 
43& 6.321\,32-07& 5.369\,82-07& 4.471\,13-07& 3.630\,70-07\\ 
44& 6.364\,38-07& 5.358\,47-07& 4.412\,53-07& 3.532\,23-07\\ 
45& 6.375\,97-07& 5.316\,08-07& 4.323\,69-07& 3.404\,73-07\\ 
46& 6.355\,31-07& 5.241\,94-07& 4.203\,97-07& 3.247\,62-07\\ 
47& 6.301\,75-07& 5.135\,43-07& 4.052\,72-07& 3.060\,22-07\\ 
48& 6.214\,75-07& 4.995\,94-07& 3.869\,24-07& 2.841\,64-07\\ 
49& 6.093\,88-07& 4.822\,84-07& 3.652\,61-07& 2.590\,53-07\\ 
50& 5.938\,69-07& 4.615\,36-07& 3.401\,53-07& 2.306\hspace{0.5cm}-07\\ 
51& 5.748\,75-07& 4.372\,43-07& 3.113\,89-07& 1.98\hspace{0.7cm}-07\\ 
52& 5.523\,47-07& 4.092\,41-07& 2.786\,1 -07& \\ 
53& 5.261\,95-07& 3.772\,61-07& 2.411\hspace{0.4cm}-07& \\ 
54& 4.962\,76-07& 3.408\,20-07& &\\ 
55& 4.623\,36-07& 2.990\hspace{0.4cm}    -07& &\\
56& 4.239\,06-07&             & &\\
\end{longtable}
\end{center}

Other transitions probabilities per second are given in Table \ref{tab:4}. 
The columns correspond to transitions between different vibrational states. 
For each $L_i$ value, the successive lines correspond to increasing values of $L_f$, 
i.e., to $L_f = L_i - 2$ for $L_i > 1$, $L_f = L_i$ for $L_i > 0$, and $L_f = L_i + 2$, respectively. 
Comparison with previous results was not possible.
 
The strongest transition from each state occurs in general towards the nearest vibrational state 
($v_f = v_i - 1$). For $v_f = v_i - 1$, in the vicinity of $L_i = 33$ and beyond, 
the  $(L_i,v_i) \rightarrow (L_i + 2,v_i-1)$ transitions 
are replaced by $(L_i + 2,v_i-1) \rightarrow (L_i,v_i)$ transitions 
because the sign of the energy difference changes 
(see the arrows in Fig.~\ref{fig:1} for the $1 \rightarrow 0$ transitions). 
These numbers are indicated in italics in Table \ref{tab:4}. 
For example, the first number in the last line for $L_i = 33$ 
corresponds to the $(35,0) \rightarrow (33,1)$ transition. 

\begin{center}
\begin{longtable}{lllllll}
\caption{Quadrupole transition probabilities per second $W$ for transitions 
between different vibrational quantum numbers $(v_i \ne v_f)$. 
The three successive lines correspond to increasing $L_f$ values, i.e.\ 
$L_f = L_i - 2$, $L_f = L_i$ and $L_f = L_i + 2$, respectively, for $L_i \ge 2$. 
Italicized numbers for $(1 \rightarrow 0)$, $(2 \rightarrow 1)$ and $(3 \rightarrow 2)$ 
mean that the initial and final states are exchanged 
(the first one is preceded in each case by a horizontal bar).} 
\label{tab:4}\\
\\[-4.9ex]
\hline
$L_i$&$(1\rightarrow\,0)$&$(2\rightarrow\,0)$&$(2\rightarrow\,1)$
     &$(3\rightarrow\,0)$&$(3\rightarrow\,1)$&$(3\rightarrow\,2)$\\
\hline
\endfirsthead
\multicolumn{7}{c}{{\tablename} \thetable{} -- Continuation}\\
\hline
$L_i$&$(1\rightarrow\,0)$&$(2\rightarrow\,0)$&$(2\rightarrow\,1)$
     &$(3\rightarrow\,0)$&$(3\rightarrow\,1)$&$(3\rightarrow\,2)$\\
\hline
\endhead
\hline
\multicolumn{7}{l}{{Continued on Next Page\ldots}}\\
\endfoot
\hline
\endlastfoot
 0&7.461\,27-08&7.384\,89-09&1.301\,15-07&5.354\,45-10&2.012\,70-08&1.693\,82-07\\ 
 \bsp                                                                      
 1&3.521\,48-08&4.090\,47-09&6.120\,20-08&3.738\,19-10&1.100\,72-08&7.940\,26-08\\ 
  &3.947\,86-08&3.459\,44-09&6.896\,88-08&2.036\,61-10&9.521\,49-09&8.994\,01-08\\ 
 \bsp                                                                      
 2&2.028\,05-08&2.703\,62-09&3.510\,24-08&2.929\,83-10&7.195\,74-09&4.535\,36-08\\ 
  &2.506\,41-08&2.918\,23-09&4.355\,33-08&2.673\,17-10&7.851\,42-09&5.649\,51-08\\ 
  &2.949\,05-08&2.250\,75-09&5.159\,38-08&9.993\,91-11&6.265\,93-09&6.737\,53-08\\ 
 \bsp                                                                      
 3&2.821\,57-08&4.089\,55-09&4.867\,92-08&4.862\,45-10&1.081\,10-08&6.268\,90-08\\ 
  &2.326\,86-08&2.718\,68-09&4.042\,32-08&2.499\,09-10&7.312\,65-09&5.242\,05-08\\ 
  &2.351\,59-08&1.530\,43-09&4.118\,62-08&4.498\,67-11&4.319\,35-09&5.383\,94-08\\ 
 \bsp                                                                      
 4&3.353\,16-08&5.255\,37-09&5.764\,14-08&6.772\,75-10&1.380\,38-08&7.395\,67-08\\ 
  &2.250\,34-08&2.641\,49-09&3.908\,07-08&2.439\,32-10&7.102\,60-09&5.066\,09-08\\ 
  &1.910\,12-08&1.030\,72-09&3.347\,91-08&1.522\,08-11&2.958\,97-09&4.379\,35-08\\ 
 \bsp                                                                      
 5&3.759\,68-08&6.344\,15-09&6.437\,06-08&8.778\,43-10&1.656\,07-08&8.225\,08-08\\ 
  &2.201\,76-08&2.599\,38-09&3.822\,12-08&2.414\,05-10&6.986\,41-09&4.952\,40-08\\ 
  &1.556\,44-08&6.702\,08-10&2.729\,11-08&2.052\,48-12&1.967\,06-09&3.570\,98-08\\ 
 \bsp                                                                      
 6&4.079\,56-08&7.386\,00-09&6.953\,87-08&1.089\,20-09&1.916\,43-08&8.844\,93-08\\ 
  &2.162\,44-08&2.570\,47-09&3.751\,99-08&2.403\,18-10&6.905\,25-09&4.858\,84-08\\ 
  &1.263\,83-08&4.112\,04-10&2.216\,18-08&5.740\,85-13&1.244\,14-09&2.899\,63-08\\ 
 \bsp                                                                      
 7&4.325\,41-08&8.379\,90-09&7.337\,19-08&1.309\,22-09&2.161\,46-08&9.285\,51-08\\ 
  &2.125\,86-08&2.547\,00-09&3.686\,35-08&2.399\,43-10&6.838\,22-09&4.770\,75-08\\ 
  &1.019\,14-08&2.306\,67-10&1.786\,58-08&7.284\,93-12&7.300\,41-10&2.336\,53-08\\ 
 \bsp                                                                      
 8&4.501\,22-08&9.313\,39-09&7.594\,90-08&1.534\,21-09&2.388\,16-08&9.558\,41-08\\ 
  &2.089\,27-08&2.525\,57-09&3.620\,48-08&2.399\,54-10&6.776\,21-09&4.682\,02-08\\ 
  &8.144\,51-09&1.117\,79-10&1.426\,82-08&1.944\,82-11&3.811\,20-10&1.864\,51-08\\ 
 \bsp                                                                      
 9&4.608\,15-08&1.016\,96-08&7.730\,22-08&1.759\,57-09&2.592\,47-08&9.669\,47-08\\ 
  &2.051\,42-08&2.504\,47-09&3.552\,20-08&2.401\,78-10&6.714\,69-09&4.589\,88-08\\ 
  &6.441\,16-09&4.103\,09-11&1.127\,27-08&3.486\,89-11&1.623\,01-10&1.471\,28-08\\ 
 \bsp                                                                      
10&4.646\,90-08&1.093\,08-08&7.745\,83-08&1.980\,23-09&2.770\,14-08&9.624\,14-08\\ 
  &2.011\,71-08&2.482\,75-09&3.480\,50-08&2.405\,07-10&6.651\,12-09&4.493\,03-08\\ 
  &5.035\,63-09&7.109\,49-12&8.800\,46-09&5.179\,80-11&4.410\,20-11&1.146\,75-08\\ 
 \bsp                                                                      
11&4.618\,86-08&1.158\,01-08&7.645\,75-08&2.191\,01-09&2.917\,28-08&9.429\,78-08\\ 
  &1.969\,86-08&2.459\,85-09&3.404\,93-08&2.408\,68-10&6.584\,00-09&4.390\,92-08\\ 
  &3.887\,87-09&4.400\,59-13&6.782\,31-09&6.887\,70-11&1.435\,85-12&8.819\,54-09\\ 
 \bsp                                                                      
12&4.526\,64-08&1.210\,32-08&7.436\,18-08&2.386\,87-09&3.030\,67-08&9.096\,67-08\\ 
  &1.925\,80-08&2.435\,38-09&3.325\,37-08&2.412\,09-10&6.512\,37-09&4.283\,44-08\\ 
  &2.961\,63-09&1.297\,61-11&5.154\,90-09&8.509\,09-11&1.307\,78-11&6.686\,30-09\\ 
 \bsp                                                                      
13&4.374\,24-08&1.248\,89-08&7.125\,79-08&2.563\,16-09&3.107\,95-08&8.638\,31-08\\ 
  &1.879\,54-08&2.409\,11-09&3.241\,87-08&2.414\,87-10&6.435\,61-09&4.170\,70-08\\ 
  &2.223\,72-09&3.806\,67-11&3.860\,00-09&9.972\,33-11&6.135\,35-11&4.991\,40-09\\ 
 \bsp                                                                      
14&4.167\,09-08&1.272\,98-08&6.725\,64-08&2.715\,82-09&3.147\,81-08&8.071\,16-08\\ 
  &1.831\,17-08&2.380\,88-09&3.154\,62-08&2.416\,68-10&6.353\,31-09&4.052\,97-08\\ 
  &1.643\,92-09&7.034\,14-11&2.844\,30-09&1.123\,11-10&1.318\,72-10&3.664\,63-09\\ 
 \bsp                                                                      
15&3.911\,91-08&1.282\,26-08&6.248\,90-08&2.841\,51-09&3.149\,95-08&7.414\,24-08\\ 
  &1.780\,82-08&2.350\,57-09&3.063\,87-08&2.417\,23-10&6.265\,23-09&3.930\,61-08\\ 
  &1.195\,04-09&1.055\,89-10&2.059\,72-09&1.225\,99-10&2.132\,39-10&2.642\,37-09\\ 
 \bsp                                                                      
16&3.616\,52-08&1.276\,80-08&5.710\,46-08&2.937\,76-09&3.115\,12-08&6.688\,50-08\\ 
  &1.728\,66-08&2.318\,13-09&2.969\,92-08&2.416\,29-10&6.171\,22-09&3.804\,07-08\\ 
  &8.529\,71-10&1.406\,25-10&1.463\,51-09&1.304\,97-10&2.967\,38-10&1.868\,03-09\\ 
 \bsp                                                                      
17&3.289\,57-08&1.257\,07-08&5.126\,39-08&3.002\,94-09&3.045\,02-08&5.916\,10-08\\ 
  &1.674\,86-08&2.283\,53-09&2.873\,13-08&2.413\,68-10&6.071\,25-09&3.673\,82-08\\ 
  &5.967\,01-10&1.731\,53-10&1.018\,38-09&1.360\,42-10&3.759\,86-10&1.292\,15-09\\ 
 \bsp                                                                      
18&2.940\,24-08&1.223\,87-08&4.513\,44-08&3.036\,36-09&2.942\,21-08&5.119\,69-08\\ 
  &1.619\,63-08&2.246\,77-09&2.773\,86-08&2.409\,26-10&5.965\,34-09&3.540\,38-08\\ 
  &4.082\,25-10&2.016\,28-10&6.923\,72-10&1.393\,60-10&4.465\,81-10&8.723\,54-10\\ 
 \bsp                                                                      
19&2.577\,98-08&1.178\,35-08&3.888\,51-08&3.038\,14-09&2.809\,99-08&4.321\,72-08\\ 
  &1.563\,17-08&2.207\,88-09&2.672\,49-08&2.402\,91-10&5.853\,59-09&3.404\,28-08\\ 
  &2.723\,93-10&2.251\,17-10&4.585\,94-10&1.406\,41-10&5.057\,60-10&5.730\,23-10\\ 
 \bsp                                                                      
20&2.212\,19-08&1.121\,86-08&3.268\,14-08&3.009\,22-09&2.652\,22-08&3.543\,75-08\\ 
  &1.505\,69-08&2.166\,92-09&2.569\,41-08&2.394\,57-10&5.736\,16-09&3.266\,04-08\\ 
  &1.766\,80-10&2.431\,75-10&2.948\,54-10&1.401\,10-10&5.520\,71-10&3.647\,94-10\\ 
 \bsp                                                                      
21&1.851\,99-08&1.055\,98-08&2.668\,08-08&2.951\,23-09&2.473\,18-08&2.805\,97-08\\ 
  &1.447\,41-08&2.123\,94-09&2.465\,01-08&2.384\,19-10&5.613\,24-09&3.126\,22-08\\ 
  &1.109\,25-10&2.557\,34-10&1.831\,81-10&1.380\,14-10&5.850\,80-10&2.239\,43-10\\ 
 \bsp                                                                      
22&1.505\,99-08&9.824\,03-09&2.102\,90-08&2.866\,37-09&2.277\,40-08&2.126\,69-08\\ 
  &1.388\,54-08&2.079\,02-09&2.359\,68-08&2.371\,77-10&5.485\,06-09&2.985\,35-08\\ 
  &6.704\,46-11&2.630\,01-10&1.093\,12-10&1.345\,99-10&6.051\,13-10&1.317\,06-10\\ 
 \bsp                                                                      
23&1.182\,08-08&9.029\,00-09&1.585\,74-08&2.757\,31-09&2.069\,56-08&1.521\,97-08\\ 
  &1.329\,29-08&2.032\,28-09&2.253\,80-08&2.357\,32-10&5.351\,90-09&2.843\,94-08\\ 
  &3.873\,38-11&2.653\,76-10&6.216\,81-11&1.301\,05-10&6.130\,33-10&7.356\,58-11\\ 
 \bsp                                                                      
24&8.873\,18-09&8.192\,67-09&1.128\,03-08&2.627\,07-09&1.854\,28-08&1.005\,46-08\\ 
  &1.269\,86-08&1.983\,80-09&2.147\,75-08&2.340\,88-10&5.214\,04-09&2.702\,52-08\\ 
  &2.118\,68-11&2.633\,84-10&3.334\,10-11&1.247\,60-10&6.100\,59-10&3.856\,46-11\\ 
 \bsp                                                                      
25&6.277\,95-09&7.332\,70-09&7.393\,77-09&2.478\,90-09&1.636\,10-08&5.881\,58-09\\ 
  &1.210\,45-08&1.933\,72-09&2.041\,88-08&2.322\,50-10&5.071\,81-09&2.561\,57-08\\ 
  &1.082\,86-11&2.576\,17-10&1.661\,45-11&1.187\,69-10&5.976\,15-10&1.865\,76-11\\ 
 \bsp                                                                      
26&4.085\,90-09&6.466\,13-09&4.274\,75-09&2.316\,15-09&1.419\,30-08&2.784\,23-09\\ 
  &1.151\,26-08&1.882\,13-09&1.936\,56-08&2.302\,27-10&4.925\,53-09&2.421\,57-08\\ 
  &5.075\,13-12&2.486\,96-10&7.529\,90-12&1.123\,16-10&5.772\,21-10&8.126\,81-12\\ 
 \bsp                                                                      
27&2.337\,34-09&5.609\,02-09&1.980\,91-09&2.142\,22-09&1.207\,89-08&8.196\,67-10\\ 
  &1.092\,46-08&1.829\,18-09&1.832\,09-08&2.280\,27-10&4.775\,54-09&2.282\,97-08\\ 
  &2.120\,59-12&2.372\,34-10&3.003\,89-12&1.055\,62-10&5.503\,99-10&3.066\,14-12\\ 
 \bsp                                                                      
28&1.062\,12-09&4.776\,18-09&5.510\,16-10&1.960\,44-09&1.005\,52-08&1.945\,02-11\\ 
  &1.034\,24-08&1.774\,98-09&1.728\,81-08&2.256\,59-10&4.622\,18-09&2.146\,18-08\\ 
  &7.552\,75-13&2.238\,16-10&9.998\,44-13&9.864\,67-11&5.186\,15-10&9.386\,87-13\\ 
 \bsp                                                                      
29&2.799\,01-10&3.981\,07-09&5.679\,29-12&1.774\,02-09&8.154\,64-09&3.909\,94-10\\ 
  &9.767\,44-09&1.719\,66-09&1.627\,01-08&2.231\,34-10&4.465\,80-09&2.011\,62-08\\ 
  &2.120\,72-13&2.089\,85-10&2.520\,89-13&9.168\,58-11&4.832\,31-10&2.060\,72-13\\ 
 \bsp                                                                      
30&7.032\,29-13&3.235\,64-09&3.485\,02-10&1.586\,00-09&6.405\,90-09&1.919\,30-09\\ 
  &9.201\,36-09&1.663\,36-09&1.526\,95-08&2.204\,62-10&4.306\,74-09&1.879\,65-08\\ 
  &4.021\,25-14&1.932\,27-10&3.927\,23-14&8.477\,62-11&4.454\,82-10&2.447\,20-14\\ 
 \bsp                                                                      
31&2.255\,30-10&2.550\,31-09&1.567\,42-09&1.399\,23-09&4.833\,62-09&4.568\,97-09\\ 
  &8.645\,55-09&1.606\,19-09&1.428\,89-08&2.176\,53-10&4.145\,34-09&1.750\,61-08\\ 
  &3.509\,86-15&1.769\,73-10&2.144\,34-15&7.799\,62-11&4.064\,60-10&6.198\,47-16\\ 
 \bsp                                                                      
32&9.471\,79-10&1.933\,95-09&3.636\,24-09&1.216\,31-09&3.458\,43-09&8.286\,36-09\\ 
  &8.101\,32-09&1.548\,27-09&1.333\,08-08&2.147\,17-10&3.981\,93-09&1.624\,83-08\\ \cline{7-7}
  &2.962\,28-17&1.605\,94-10&1.306\,19-18&7.140\,81-11&3.671\,06-10&{\it 6.430\,53-22}\\ 
 \bsp                                                                      
33&2.151\,10-09&1.394\,00-09&6.516\,22-09&1.039\,62-09&2.297\,18-09&1.300\,18-08\\ 
  &7.569\,86-09&1.489\,72-09&1.239\,72-08&2.116\,64-10&3.816\,82-09&1.502\,59-08\\ \cline{2-2}\cline{4-4}
  &{\it 1.611\,20-18}&1.443\,99-10&{\it 8.424\,15-17}&6.506\,00-11&3.282\,16-10&{\it 9.779\,29-16}\\ 
 \bsp                                                                      
34&3.816\,28-09&9.364\,63-10&1.015\,77-08&8.712\,85-10&1.363\,18-09&1.863\,16-08\\ 
  &7.052\,25-09&1.430\,66-09&1.149\,00-08&2.085\,02-10&3.650\,33-09&1.384\,15-08\\ 
  &{\it 1.139\,31-15}&1.286\,46-10&{\it 6.745\,28-15}&5.898\,76-11&2.904\,43-10&{\it 2.587\,17-14}\\ 
 \bsp                                                                      
35&5.916\,19-09&5.661\,05-10&1.450\,18-08&7.132\,37-10&6.665\,25-10&2.508\,06-08\\ 
  &6.549\,48-09&1.371\,19-09&1.061\,11-08&2.052\,39-10&3.482\,74-09&1.269\,75-08\\ 
  &{\it 1.699\,91-14}&1.135\,40-10&{\it 6.503\,38-14}&5.321\,66-11&2.543\,11-10&{\it 1.791\,55-13}\\ 
 \bsp                                                                      
36&8.419\,63-09&2.865\,40-10&1.948\,18-08&5.671\,93-10&2.144\,55-10&3.224\,34-08\\ 
  &6.062\,41-09&1.311\,41-09&9.762\,01-09&2.018\,80-10&3.314\,35-09&1.159\,59-08\\ 
  &{\it 9.475\,42-14}&9.923\,81-11&{\it 3.006\,77-13}&4.776\,40-11&2.202\,22-10&{\it 7.070\,43-13}\\ 
 \bsp                                                                      
37&1.129\,16-08&1.004\,02-10&2.502\,46-08&4.347\,06-10&1.174\,03-11&4.000\,66-08\\ 
  &5.591\,82-09&1.251\,41-09&8.944\,05-09&1.984\,29-10&3.145\,40-09&1.053\,86-08\\ 
  &{\it 3.331\,38-13}&8.585\,86-11&{\it 9.536\,97-13}&4.263\,96-11&1.884\,75-10&{\it 2.045\,74-12}\\ 
 \bsp                                                                      
38&1.449\,42-08&9.504\,00-12&3.105\,20-08&3.172\,01-10&6.109\,78-11&4.824\,99-08\\ 
  &5.138\,37-09&1.191\,30-09&8.158\,44-09&1.948\,89-10&2.976\,16-09&9.527\,14-09\\ 
  &{\it 8.962\,46-13}&7.348\,19-11&{\it 2.405\,64-12}&3.784\,75-11&1.592\,73-10&{\it 4.863\,40-12}\\ 
 \bsp                                                                      
39&1.798\,70-08&1.501\,27-11&3.748\,19-08&2.160\,20-10&3.636\,02-10&5.684\,75-08\\ 
  &4.702\,65-09&1.131\,15-09&7.406\,20-09&1.912\,59-10&2.806\,87-09&8.562\,88-09\\ 
  &{\it 2.025\,12-12}&6.215\,76-11&{\it 5.204\,46-12}&3.338\,74-11&1.327\,39-10&{\it 1.010\,23-11}\\ 
 \bsp                                                                      
40&2.172\,82-08&1.176\,26-10&4.422\,92-08&1.324\,80-10&9.191\,03-10&6.566\,87-08\\ 
  &4.285\,13-09&1.071\,03-09&6.688\,20-09&1.875\,35-10&2.637\,74-09&7.646\,95-09\\ 
  &{\it 4.049\,06-12}&5.190\,88-11&{\it 1.009\,13-11}&2.925\,52-11&1.089\,27-10&{\it 1.902\,85-11}\\ 
 \bsp                                                                      
41&2.567\,48-08&3.177\,61-10&5.120\,63-08&6.794\,19-11&1.726\,63-09&7.457\,85-08\\ 
  &3.886\,23-09&1.011\,03-09&6.005\,13-09&1.837\,10-10&2.468\,99-09&6.780\,27-09\\ 
  &{\it 7.398\,83-12}&4.273\,64-11&{\it 1.803\,42-11}&2.544\,42-11&8.783\,41-11&{\it 3.329\,82-11}\\ 
 \bsp                                                                      
42&2.978\,34-08&6.157\,51-10&5.832\,39-08&2.389\,56-11&2.784\,81-09&8.343\,71-08\\ 
  &3.506\,27-09&9.512\,02-10&5.357\,56-09&1.797\,72-10&2.300\,79-09&5.963\,54-09\\ 
  &{\it 1.262\,40-11}&3.462\,31-11&{\it 3.027\,44-11}&2.194\,56-11&6.940\,82-11&{\it 5.505\,31-11}\\ 
 \bsp                                                                      
43&3.401\,01-08&1.012\,05-09&6.549\,09-08&2.062\,35-12&4.092\,21-09&9.209\,92-08\\ 
  &3.145\,50-09&8.916\,09-10&4.745\,92-09&1.757\,05-10&2.133\,34-09&5.197\,30-09\\ 
  &{\it 2.041\,79-11}&2.753\,65-11&{\it 4.839\,39-11}&1.874\,94-11&5.355\,93-11&{\it 8.706\,31-11}\\ 
 \bsp                                                                      
44&3.831\,15-08&1.507\,46-09&7.261\,45-08&4.531\,33-12&5.647\,71-09&1.004\,13-07\\ 
  &2.804\,10-09&8.323\,02-10&4.170\,50-09&1.714\,83-10&1.966\,78-09&4.481\,93-09\\ 
  &{\it 3.165\,34-11}&2.143\,26-11&{\it 7.441\,71-11}&1.584\,44-11&4.016\,57-11&{\it 1.329\,46-10}\\ 
 \bsp                                                                      
45&4.264\,40-08&2.103\,40-09&7.959\,94-08&3.393\,63-11&7.450\,75-09&1.082\,15-07\\ 
  &2.482\,19-09&7.733\,28-10&3.631\,49-09&1.670\,74-10&1.801\,24-09&3.817\,61-09\\ 
  &{\it 4.743\,79-11}&1.625\,82-11&{\it 1.109\,66-10}&1.321\,92-11&2.908\,09-11&{\it 1.975\,15-10}\\ 
 \bsp                                                                      
46&4.696\,48-08&2.802\,13-09&8.634\,65-08&9.369\,47-11&9.501\,41-09&1.153\,27-07\\ 
  &2.179\,82-09&7.147\,22-10&3.128\,97-09&1.624\,31-10&1.636\,82-09&3.204\,43-09\\ 
  &{\it 6.919\,50-11}&1.195\,28-11&{\it 1.615\,01-10}&1.086\,19-11&2.013\,85-11&{\it 2.873\,44-10}\\ 
 \bsp                                                                      
47&5.123\,08-08&3.607\,18-09&9.275\,08-08&1.883\,36-10&1.180\,02-08&1.215\,46-07\\ 
  &1.896\,99-09&6.565\,12-10&2.662\,92-09&1.574\,87-10&1.473\,56-09&2.642\,31-09\\ 
  &{\it 9.879\,18-11}&8.450\,64-12&{\it 2.307\,12-10}&8.760\,65-12&1.315\,56-11&{\it 4.117\,29-10}\\ 
 \bsp                                                                      
48&5.539\,92-08&4.523\,64-09&9.869\,73-08&3.239\,54-10&1.434\,67-08&1.266\,30-07\\ 
  &1.633\,67-09&5.987\,07-10&2.233\,22-09&1.521\,47-10&1.311\,42-09&2.131\,07-09\\ 
  &{\it 1.387\,40-10}&5.681\,54-12&{\it 3.251\,59-10}&6.903\,90-12&7.934\,92-12&{\it 5.844\,03-10}\\ 
 \bsp                                                                      
49&5.942\,57-08&5.558\,60-09&1.040\,54-07&5.088\,20-10&1.713\,62-08&1.302\,65-07\\ 
  &1.389\,76-09&5.412\,95-10&1.839\,69-09&1.462\,66-10&1.150\,21-09&1.670\,42-09\\ 
  &{\it 1.925\,22-10}&3.571\,82-12&{\it 4.543\,96-10}&5.280\,29-12&4.264\,14-12&{\it 8.268\,44-10}\\ 
 \bsp                                                                      
50&6.326\,40-08&6.721\,46-09&1.086\,61-07&7.541\,0\hspace{0.2cm}-10
  &2.015\,0\hspace{0.2cm}-08&1.320\,1\hspace{0.2cm}-07\\ 
  &1.165\,13-09&4.842\,28-10&1.482\,07-09&1.396\,0\hspace{0.2cm}-10
  &9.894\,4\hspace{0.2cm}-10&1.260\,0\hspace{0.2cm}-09\\ 
  &{\it 2.651\,49-10}&2.044\,26-12&{\it 6.330\,46-10}&3.878\,8\hspace{0.2cm}-12
  &1.913\hspace{0.5cm}-12&{\it 1.175\,3\hspace{0.2cm}-09}\\ 
 \bsp                                                                      
51&6.686\,25-08&8.023\,74-09&1.123\,02-07
  &1.074\hspace{0.5cm}-09&2.333\hspace{0.5cm}-08&1.312\hspace{0.5cm}-07\\ 
  &9.596\,11-10&4.273\,85-10&1.160\,03-09
  &1.317\hspace{0.5cm}-10&8.28\hspace{0.7cm}-10 &8.99\hspace{0.7cm}-10\\ 
  &{\it 3.641\,53-10}&1.017\,29-12&{\it 8.849\,45-10}
  &2.689\hspace{0.5cm}-12&6.24\hspace{0.7cm}-13 &{\it 1.69\hspace{0.7cm}-09}\\ 
 \bsp                                                                      
52&7.015\,89-08&9.476\,8\hspace{0.2cm}-09&1.146\,6\hspace{0.2cm}-07&&&\\ 
  &7.730\,11-10&3.705\,1\hspace{0.2cm}-10&8.731\,4\hspace{0.2cm}-10&&&\\ 
  &{\it 5.014\,92-10}&4.030\,2\hspace{0.2cm}-13&{\it 1.252\,4\hspace{0.2cm}-09}&&&\\ 
 \bsp                                                                      
53&7.306\,88-08&1.108\hspace{0.5cm}-08&1.151\,6\hspace{0.2cm}-07&&&\\ 
  &6.050\,95-10&3.130\hspace{0.5cm}-10&6.2\hspace{0.9cm}-10&&&\\ 
  &{\it 6.975\,09-10}&1.039\hspace{0.5cm}-13&{\it 1.82\hspace{0.7cm}-09}&&&\\ 
 \bsp                                                                      
54&7.545\,71-08&&&&&\\ 
  &4.555\,87-10&&&&&\\ 
  &{\it 9.903\hspace{0.5cm}-10}&&&&&\\ 
\end{longtable}
\end{center}

\begin{figure}[hbt]
\setlength{\unitlength}{1mm}
\begin{picture}(140,50) (-20,10) 
\put(0,0){\mbox{\scalebox{1.2}{\includegraphics{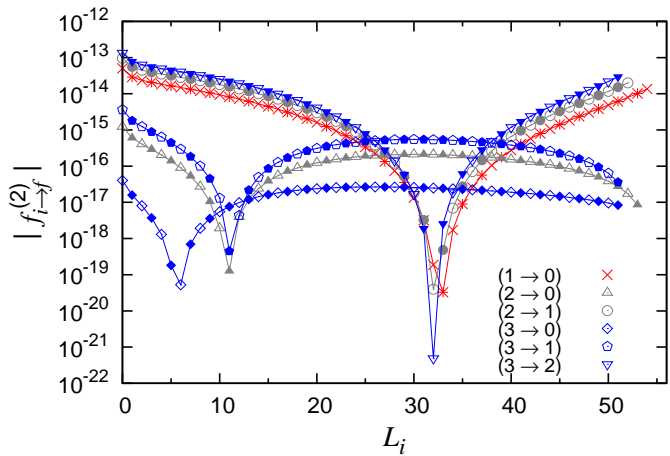}}}}
\end{picture} \\
\caption{Oscillator strengths for $L_f = L_i + 2$ transitions.}
\label{fig:2}
\end{figure}
\begin{figure}[hbt]
\setlength{\unitlength}{1mm}
\begin{picture}(140,55) (-20,10) 
\put(0,0){\mbox{\scalebox{1.2}{\includegraphics{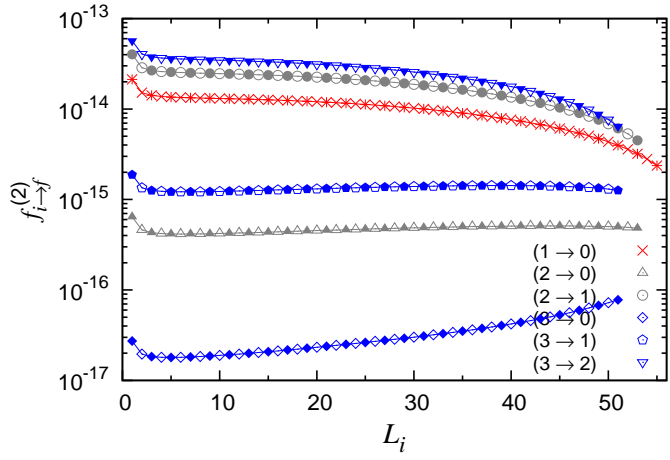}}}}
\end{picture} \\
\caption{Oscillator strengths for $L_f = L_i$ transitions.}
\label{fig:3}
\end{figure}
\begin{figure}[hbt]
\setlength{\unitlength}{1mm}
\begin{picture}(140,55) (-20,10) 
\put(0,0){\mbox{\scalebox{1.2}{\includegraphics{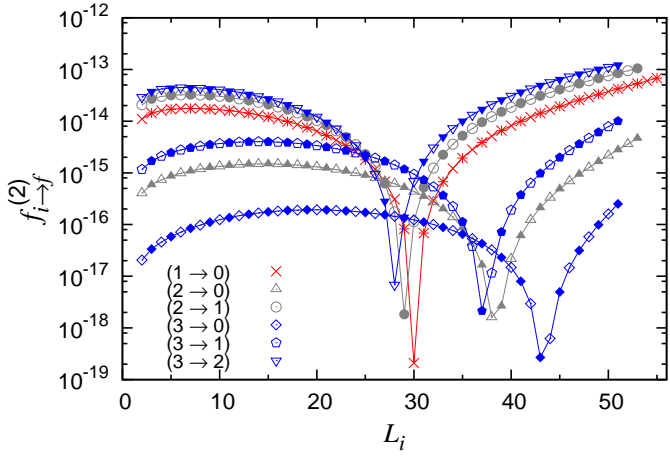}}}}
\end{picture} \\
\caption{Oscillator strengths for $L_f = L_i - 2$ transitions.}
\label{fig:4}
\end{figure}

Figs.~\ref{fig:2}, \ref{fig:3} and \ref{fig:4} summarize the oscillator strengths for most of the considered transitions. For $\Delta L = L_f - L_i = 2$ (Fig.~\ref{fig:2}) the oscillator strength has a strong variation along the band varying also with $\Delta v = v_i - v_f$. 
The strong minima for $\Delta v = 1$ around $L_i = 33$ are due to the change of sign of the energy difference (see Fig.~\ref{fig:1}). 
Beyond that value, the initial state is lower than the final $v_f < v_i$ state and 
the strengths are negative. 
Otherwise the strengths slowly increase with the vibrational excitation. 
The $\Delta v = 2$ strengths are smaller by more than an order of magnitude. 
The minimum occurring around $L_i = 11$ is here due to a change of sign of the matrix element 
appearing in \eref{2.3}. 
The $\Delta v = 3$ strengths are smaller by more than an order of magnitude 
than the $\Delta v = 2$ strengths. 

The $\Delta L = 0$ oscillator strengths presented in Fig.~\ref{fig:3} do not vary much along the bands. 
They slowly decrease for $\Delta v = 1$ and slowly increase for $\Delta v = 3$. 
They are remarkably flat for $\Delta v = 2$. 

The $\Delta L = -2$ strengths presented in Fig.~\ref{fig:4} behave similarly 
to the $\Delta L = 2$ strengths. 
Minima take place around $L_i = 29$, $37$ and $43$ for $\Delta v = 1$, $2$ and $3$, respectively.
These minima thus occur now at increasing $L_i$ values with increasing $\Delta v$ 
and are all due to a change of sign of the matrix element. 

\begin{figure}[hbt]
\setlength{\unitlength}{1mm}
\begin{picture}(140,60) (-20,7) 
\put(0,0){\mbox{\scalebox{1.2}{\includegraphics{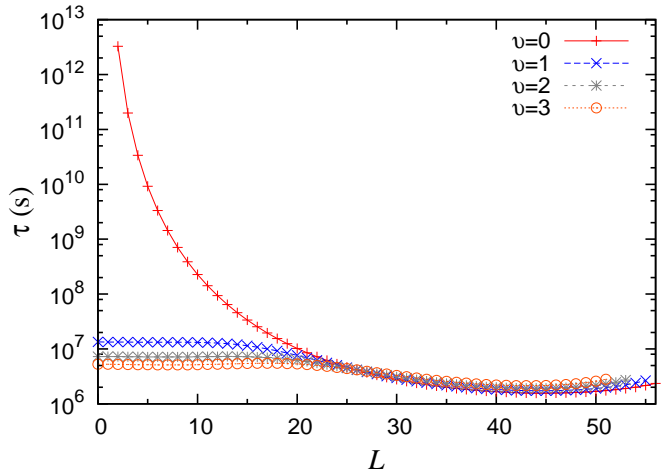}}}}
\end{picture} \\
\caption{Lifetimes $\tau$ in the first four rotational bands ($v = 0-3$).}
\label{fig:5}
\end{figure}

The lifetimes can be calculated as
\beq
\tau = \bigg( \sum_{E_f < E_i} W_{i \rightarrow f} \bigg)^{-1}.
\eeqn{6.1} 
Using the transitions probabilities of Table \ref{tab:4} these are, for states with $L=0$ $1.340\,255\times10^{7}$, $7.272\,729\times10^{6}$ and $5.261\,929\times10^{6}$ for $v=1$, $v=2$ and $v=3$, respectively. Lifetimes for the calculated states are displayed in Fig.~\ref{fig:5}. 

\section{Conclusion}
\label{sec:conc}
Applying the Langrange-mesh method, the three-body Schr\"odinger equation for the deuterium molecular ion was solved with high precision in perimetric coordinates. Four vibrational states $v=0-3$ and $L$ form $0$ to $56$ were considered. With the same number of mesh points in the $x$ and $y$ coordinates $N=40$ and $N_z=20$ for all the sates considered, energies values are presented with 13 significant digits for the zero vibrational band until 9 significant digits for the third  vibrational band.

With the wave functions provided by the Lagrange-mesh method, a simple calculation gives the quadrupole strengths and transition probabilities per time unit. Extensive tables and some figures are presented with six significant digits. 

For high values of the total angular momentum $L$ and  $\Delta v= 1$, the initial and final states  are interchange: $L=33$, $33$ and $32$ for $v_i=0 \rightarrow v_f=1$, $v_i=1 \rightarrow v_f=2$ and $v_i=2 \rightarrow v_f=3$, respectively. In comparison with the hydrogen molecular ion~\cite{OB12T} this values are $L=23$, $23$ and $22$ for the same change in the vibrational bands. Also in a consistent way, quadrupole transitions for the deuteron molecular ion are less favorable that those for the hydrogen molecular ion, except for some values of the total angular momentum. As a consequence, lifetimes for D$_2^+$ are greater than those for H$_2^+$. The difference is around one order of magnitude except for the states in the ground-state rotational band, roughly $10^7$ and $10^6$, respectively.
\ack
I indebted to D. Baye for initiating me into the subject and A. Turbiner for suggesting the problem and for their interest. I would like to thanks the F.R.S.-FNRS (Belgium) for a postdoctoral grant. 
\section*{References}
%

%
\end{document}